\documentclass[shortnote]{jpsj2}

\title{%
Duality and Multicritical Point of Two-Dimensional Spin Glasses
}

\author{%
Hidetoshi \textsc{Nishimori} and
Koji \textsc{Nemoto}$^{1}$
}

\inst{%
Department of Physics, Tokyo Institute of Technology,
Oh-okayama, Meguro-ku, Tokyo 152-8551 \\
$^1$Division of Physics, Hokkaido University, Sapporo 060-0810
}

\recdate{\today}

\abst{%
}

\kword{%
spin glass, multicritical point, duality
}

\begin{document}
\sloppy
\maketitle

Determination of the precise location of the multicritical point and
phase boundary is a target of active current research in the
theory of spin glasses.\cite{Aarao,Singh-Adler,Ozeki-Ito,
Honecker-Picco,Mertz-Chalker,
Jacobsen-Picco01,Cho-Fisher,Gruzberg,Read-Ludwig,
Senthil-Fisher,Sorensen}
In this short note we develop a duality argument
to predict the location of the multicritical point and the shape of
the phase boundary in models of spin glasses on the square lattice.

The first system we treat is a random $Z_q$ model with gauge symmetry
which includes the $\pm J$ Ising model and
the Potts gauge glass.
Following the notation of Wu and Wang,~\cite{Wu-Wang76} the partition
function is
 \begin{equation}
  Z=\sum_{\{ \xi_i\} } \exp\left\{ \sum_{\langle ij\rangle}
      V(\xi_i -\xi_j +J_{ij}) \right\}.
  \end{equation}
Here, $\xi_i(=0,1,\cdots,q-1)$ is the $q$-state spin variable,
$J_{ij}(=0,1,\cdots,q-1)$ is the quenched randomness, and
$V(\cdot )$ is a periodic function with period $q$.
The sum in the exponent runs over neighbouring sites on
the square lattice.
It is straightforward to generalize the duality transformation
in the Wu-Wang formalism~\cite{Wu-Wang76} to the random model,
and the result is
 \begin{eqnarray}
   Z &=& \sum_{\{\eta\}} {}' \exp\left\{
     \sum_{\langle ij\rangle} V(\eta_{ij} +J_{ij})\right\} 
     \label{Z1}\\
    & =& q^N \sum_{\{\lambda\}}{}' \exp\left\{\sum_{\langle ij\rangle}
       \tilde{V}(\lambda_{ij})
       +\frac{2\pi {\rm i}}{q} \lambda_{ij} J_{ij} \right\},
     \label{Z2}
  \end{eqnarray}
where $\eta_{ij}$ denotes $\xi_i -\xi_j$ and $\lambda_{ij}$ is
the bond variable on the dual lattice corresponding to $\eta_{ij}$
and ${\rm e}^{\tilde V}$ is the Fourier transform of ${\rm e}^{V}$.
The prime in the sum indicates that the bond variables 
$\eta_{ij}$ and $\lambda_{ij}$ are constrained
such that their sums over a closed loop vanish.~\cite{Wu-Wang76}
Thus the duality transformation consists in replacing the exponential
in (\ref{Z1}) with that in (\ref{Z2}):
\begin{equation}
{\rm  e}^{V(\eta +J)}\leftrightarrow {\rm  e}^{\tilde{V}(\lambda )
   +2\pi {\rm  i} \lambda J/q}.
 \label{duality}
\end{equation}

Average over quenched randomness is dealt with by the replica method.
A generalization of the existing argument~\cite{Nishimori79}
suggests that it is convenient to consider the following quantity
 \begin{equation}
  \alpha =\frac{\displaystyle
   \sum_{l=0}^{q-1} p_l \left( \sum_{\eta=0}^{q-1}
  {\rm  e}^{V(\eta +l)}\right)^n}{\displaystyle
    \sum_{l=0}^{q-1} p_l \, {\rm  e}^{nV(l)}},
 \end{equation}
where $p_l$ is the probability that $J_{ij}$ takes the value $l$.
It is not difficult to check that the duality transformation
(\ref{duality}) changes $\alpha$ into $\tilde{\alpha}=q^n/\alpha$.
The subspace defined by $\alpha =q^{n/2}$ in the parameter space
$(p_1, p_2,\cdots,T,\cdots )$
is therefore left invariant by the duality transformation:
A point in this subspace is transformed into another point
in the same subspace although the latter point
does not necessarily have real values of Boltzmann weights.
In general it is possible to consider complex values of Boltzmann
weights (or parameters such as the temperature);
the subspace $\alpha =q^{n/2}$ embedded in the
complex-valued parameter space maps onto itself by the duality
transformation (\ref{duality}).
We shall consider the cross section of this subspace with
the phase diagram with real parameters, which we shall call
the invariant subspace in the following.

One should be careful in identifying the invariant subspace with
a critical surface.~\cite{Aharony-Stephen80}
We discuss this problem later; let us see here what
the invariant subspace looks like in the quenched system,
which is obtained by taking the limit $n\to 0$ of the relation 
$\alpha =q^{n/2}$:
 \begin{equation}
    \sum_{l=0}^{q-1} p_l \log \left(
      \sum_{\eta =0}^{q-1} {\rm  e}^{V(\eta +l)-V(l)} 
          \right)=\frac{1}{2}\log q.
    \label{invsurf}
 \end{equation}

\begin{figure}[hb]
  \begin{center}
  \includegraphics[width=.28\linewidth]{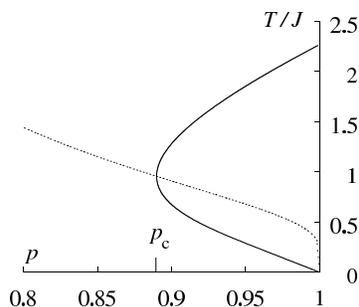}
  \end{center}
  \caption{Invariant subspace of the $\pm J$ Ising model
   is the curve drawn by the solid line.
   Shown dashed is the Nishimori line.}
  \label{fig:1}
\end{figure}

We now apply (\ref{invsurf}) to a few examples and
discuss the consequences.
The first case is the $\pm J$ Ising model ($q=2$) with $V(0)=\beta J,
V(1)=-\beta J$ and $p_0=p, p_1=1-p$.
The invariant subspace (\ref{invsurf}) then reads
  \begin{equation}
  p\log (1+{\rm  e}^{-2\beta J}) +(1-p) \log (1+{\rm  e}^{2 \beta J})
       =\frac{1}{2}\log 2.
 \label{solution-IsingJ}
 \end{equation}
This curve is plotted in Fig. \ref{fig:1} in the $p$-$T$ phase diagram.
It is observed (and can be confirmed analytically)
that the curve reaches the point with minimum $p$
on the Nishimori line~\cite{Nishimori81} ${\rm  e}^{-2\beta J}=(1-p)/p$,
from which and (\ref{solution-IsingJ}),
the minimum point $p_{\rm c}$ is found to satisfy
 \begin{equation}
   -p_{\rm c}\log p_{\rm c}-(1-p_{\rm c})\log (1-p_{\rm c})
  =\frac{\log 2}{2},
 \end{equation}
or $p_{\rm c}=0.889972$.  This coincides with
numerical estimates of $p$ at the multicritical point with high
precision: 0.8905(5),\cite{Aarao} 0.886(3),\cite{Singh-Adler}
0.8872(8),\cite{Ozeki-Ito} 0.8906(2)\cite{Honecker-Picco}
and 0.8907(2).\cite{Mertz-Chalker}

The next example is the $q$-state Potts gauge glass with
$V(0)=\beta J, V(1)=V(2)=\cdots =V(q-1)=0$
and $p_0=1-(q-1)p, p_1=\cdots =p_{q-1}=p$, where
we have followed the notation of Jacobsen and Picco.\cite{Jacobsen-Picco01}
The invariant subspace (\ref{invsurf}) is, in this case,
 \begin{eqnarray}
  &&\{ 1-(q-1)p\} \log \{1+(q-1){\rm  e}^{-\beta J}\}
    \nonumber\\
    &&+(q-1) p \log (q-1 +{\rm  e}^{\beta J}) =\frac{\log q}{2}.
     \label{solution-Potts}
 \end{eqnarray}
This formula reduces to the Ising counterpart (\ref{solution-IsingJ})
when $q=2$ under appropriate changes of energy scale and
notation (exchange of $p$ and $1-p$).
The curve specified by (\ref{solution-Potts}) reaches the extremum
point in the $p$-$T$ phase diagram again on the Nishimori line
$p=1/({\rm  e}^{\beta J}+q-1)$.\cite{Nishimori-Stephen83}
In particular, for $q=3$, this extremum point $p_{\rm c}$ satisfies
 \begin{equation}
 -(1-2p_{\rm c})\log (1-2p_{\rm c}) -2 p_{\rm c} \log p_{\rm c}
     =\frac{\log 3}{2},
  \end{equation}
whose numerical value $p_{\rm c}=0.0797308$ coincides very well
with the location of the multicritical point,
$0.079$-$0.080$.\cite{Jacobsen-Picco01}

It is possible to apply the same argument to the random Ising model
with a general distribution function $P(J_{ij})$ of
exchange interactions.\cite{Nishimori79}
The result for the invariant subspace is
  \begin{equation}
    \int_{-\infty}^{\infty} {\rm  d}J_{ij}\,
      P(J_{ij})\,\log (1+{\rm  e}^{-2\beta J_{ij}})
  =\frac{\log 2}{2}.
   \label{general}
  \end{equation}
This formula applied to the Gaussian model with
$P(J_{ij})\propto{\rm  e}^{-(J_{ij}-J_0)^2/2J^2}$
represents a curve in the phase diagram similar to
that of the $\pm J$ model
in Fig. \ref{fig:1} and again has the smallest $J_0/J$ on the Nishimori
line $\beta J^2 =J_0$.~\cite{Nishimori81}
This point is at $J_0/J=1.02177$ which is very close to
the multicritical point evaluated numerically
(Horie, Nemoto, Hukushima and Ozeki, private communication).

What do all these results mean?
One possibility is that the exact locations of multicritical points
have been derived for the above models.
The present argument using the subspace invariant under duality
is, in general, not guaranteed to give the exact phase boundary
or multicritical point.
Nevertheless we have several reasons to conjecture that our result
may be exact, in particular for the multicritical point.

The first reason is that the invariant subspace $\alpha =q^{n/2}$
coincides with the exact phase boundary for the $\pm J$ Ising
model in the case of $n=1$ and $n=2$.~\cite{Georges-etal87}
It should be noticed that this coincidence
exists only above the multicritical point in the phase diagram
for the $n=2$ case, which may be related to the strange shape
of the curve below the multicritical point in Fig. \ref{fig:1}.
The second reason is the impressive agreement with numerical estimates
explained above,
which seems to us beyond a simple coincidence.
The third evidence is the fact that the Nishimori line appears
naturally from the present duality formalism.
It is well established that the multicritical point is located on the
Nishimori line for models with gauge symmetry.~\cite{Nishimori01}
The natural emergence of the Nishimori line from arguments
without explicit use of gauge transformation strongly suggests that
something deep may be hidden behind the scene.

There are of course several problems that deserve special caution.
As pointed out by Aharony and Stephen,\cite{Aharony-Stephen80}
duality does not yield fixed points of the transformation for random
models whereas fixed points of duality are often
identical to critical points in non-random systems.
Aharony and Stephen thus argued that
duality in random systems, unable to identify fixed points, is
not a useful tool of analysis.
We are suggesting here that an invariant subspace, if not fixed points,
may be of some use in random systems.

Another problem is whether or not the whole invariant subspace
(\ref{invsurf}) or (\ref{general}) coincides with the phase boundary.
It is dangerous to accept this identification at least
below the multicritical point in the phase diagram because of
the $\pm J$ Ising model with $n=2$ as mentioned above.
The origin of disagreement of our result (\ref{solution-IsingJ})
near $p=1$ with the perturbative calculation of Domany\cite{Domany79}
should also be clarified in future investigations.


\end{document}